\documentstyle[preprint,eqsecnum,aps]{revtex}

\begin{document}
\draft
\title{Accurate
Cosmological Parameters and Supersymmetric\\ 
Particle Properties}
\author{Pran Nath$^1$ and R. Arnowitt$^2$}
\address
{$^1$Department of Physics, Northeastern University, Boston, MA 02115,\\
$^2$Center for Theoretical Physics, Department of Physics,\\
Texas A\&M University, College Station, TX 77843-4242 
}
\date{\today}
\maketitle
\begin{abstract}
Future sattelite, balloon and ground based experiments will give precision  
determinations  of the basic cosmological parameters and hence determine 
the amount of cold dark matter in the universe accurately. 
We consider here two cosmological models, the $\nu$CDM model and the 
$\Lambda$CDM model, and examine within the framework of supergravity 
grand unification the effect this will have for these models on supersymmetry 
searches at accelerators.
In the former example the gluino (neutralino) mass has an upper bound 
of about 720(100) $GeV$ and gaps (forbidden regions) may develop at 
lower energies.
In the latter case the upper bound occurs at gluino (neutralino) mass of 
about 520(70) $GeV$ with the squarks and selectron becoming light when 
gluino (neutralino) masses are greater than 420(55) $GeV$. 
Both models are sensitive to non-universal soft breaking masses, and show 
a correlation between large (small) dark matter detector event rates and low 
(high) $b\rightarrow s+\gamma$ branching ratio.
\end{abstract}
\pacs{12.60.Jv, 98.80.C, 95.35.+d}

\narrowtext

While  particle physics and cosmology have always interacted in an important 
way, the development of supersymmetry (SUSY) has greatly deepened this 
relation.
Thus SUSY models with R-parity invariance predict the existence of cold dark 
matter  (CDM) which may make up between 40\% and 80\% of all the matter of 
the universe.
SUSY allows one to construct theories that reach in energy up to the GUT scale 
($M_G\cong 10^{16}\ GeV$) and backwards in time to the very early universe 
and are consistent with all known data. 
In this letter we study the effects that accurate determinations of the 
cosmological parameters will have on the SUSY spectrum.
The new sattelites, MAP and Planck \cite{1} 
and the many ground based and balloon experiments, 
will measure the basic cosmological parameters, the Hubble 
constant $H$, the cosmological constant $\Lambda$, the amount of matter in the 
universe $\Omega= \rho/\rho_c$ (where $\rho$ is the density of matter and 
$\rho_c= 3H^2/8\pi G_N$, $G_N=$ Newton's constant) etc. at the level of a few 
percent.
While at present there are many possible cosmological models that fit the 
current data leading to the wide window 
$0.1 \alt
\Omega_{CDM} h^2 \alt 0.4$   
(where $h= (H/100)Mpc^{-1} s^{-1})$, 
the new experiments will fix this quantity very 
accurately. 
This will then greatly restrict the allowed SUSY parameter space, which in turn 
will influence SUSY predictions at accelerators.

In this letter we consider two possibilities, ({\it i}) the $\nu$CDM model (with CDM, neutrino hot dark matter 
(HDM) and baryonic (B) dark matter) and ({\it ii}) the $\Lambda$CDM model 
(with CDM, a cosmological constant $\Lambda$ and B dark matter).
In order to analyse these, it is necessary to have a fixed SUSY dynamics, and we use 
here supergravity grand unification models with R-parity where gravity mediates 
the SUSY breaking in a hidden sector at a scale 
$\agt M_G$ 
and gravity as the messenger of this breaking to the physical sector \cite{2}. 
We assume here universal gaugino masses at the GUT scale, and also universal 
scalar masses $m_0$ for the first two generations of squarks and sleptons at $M_G$ 
(to suppress flavor changing neutral currents).
We parameterize the Higgs $H_{1,2}$ and third generation scalar masses at $M_G$ 
by 
$m_{H_{1,2}}^2=m_0^2(1+\delta_{1,2})$,
$m_{q_L}^2=m_0^2(1+\delta_3)$,    
$m_{u_R}^2=m_0^2(1+\delta_4)$,  
$m_{e_R}^2=m_0^2(1+\delta_5)$,  
$m_{d_R}^2=m_0^2(1+\delta_6)$,  
and $m_{l_L}^2=m_0^2(1+\delta_7)$,  
where $q_L\equiv (\tilde{u}_L, \tilde{d}_L)$  is the $L$ squark doublet, $u_R$ the 
right up(top)-squark singlet etc. 
In addition, there are third generation cubic soft breaking 
parameters $A_{0t}$, $A_{0b}$, $A_{0\tau}$ for the $t$, $b$ and $\tau$ particles. 
In the following we restrict the parameters to the range 
$m_0, m_{\tilde{g}} \le 1\ TeV$,  $|A_t/m_0| \le 7$, 
 $|\delta_i| \le 1$,  $\tan\beta \le 25$, 
where $m_{\tilde{g}}$ is the gluino mass, $\tan\beta = \langle H_2 \rangle /
\langle H_1 \rangle$,  $\langle H_{1,2} \rangle$ gives masses to the $(d,u)$ 
quarks and $A_t$ is the $t$-quark parameter at the electroweak scale. 
The limit on $\tan\beta$ implies that $\delta_{5,6,7}$ and $A_{b,\tau}$ make only 
small contributions and so we set these to zero in the following.

Radiative breaking of 
$SU(2)\times U(1)$ at the electroweak scale 
\cite{3,4}
determines the Higgs mixing parameter $\mu$ 
($W_{\mu}=\mu H_1H_2$) at the $Z$ boson mass $M_Z$ to be \cite{5}
\widetext
\begin{eqnarray}
\mu^2=&&{t^2\over t^2-1}\biggl[\biggl\{ {1-3D_0\over 2}+{1\over t^2} \biggr\}+
   \biggl\{ {1-D_0\over 2}(\delta_3+\delta_4) - {1+D_0\over 2}\delta_2 
  + {1\over t^2}\delta_1 \biggr\} \biggr] m_0^2\nonumber\\
&&+ {t^2\over t^2-1}
\biggl[{1\over 2}(1-D_0){A_R^2\over D_0} + C_{\tilde{g}}m_{\tilde{g}}^2\biggr] 
- {1\over 2}M_Z^2 + {1\over 22}{t^2+1\over t^2-1}
\biggl(1+{\alpha_1\over \alpha_G}\biggr)S_0 + loop\ corrections
\eqnum{1} 
\label{eq:2}
\end{eqnarray}
\narrowtext
\noindent
where $t\equiv \tan\beta$, $D_0\cong 1-(m_t/200\sin\beta)^2$, 
$A_R\cong A_t-0.613m_{\tilde{g}}$, $S_0= Tr Ym^2$ ($Y=$ hypercharge, 
$m^2=$ masses at $M_G$), $\alpha_1(M_Z)$ is the $U(1)$ gauge coupling constant, 
$\alpha_G \cong 1/24$ is the GUT scale coupling constant and $C_{\tilde{g}}$ is 
given in \cite{4}.
In Eq.\ (\ref{eq:2}), $D_0$ vanishes at the $t$- quark Landau pole and 
$A_R$ is the residue at 
the pole. 
Thus $D_0$ is generally small ($D_0\le 0.23$ for $m_t= 175\ GeV$) and hence the 
non-universal effects are governed approximately by the combination  
$\delta_2-(\delta_3+\delta_4)$.

For much of the parameter space, one finds that $\mu^2 \gg M_Z^2$ and ``scaling" 
relations  hold 
among the neutralinos ($\chi_i^0$, $i=1,...4$) and the charginos ($\chi_i^{\pm}$, 
$i=1,2$) \cite{6}:
$2m_{\chi_1^0} \cong m_{\chi_1^{\pm}} \cong m_{\chi_2^0} 
\cong ({1\over 3} - {1\over 4}) m_{\tilde{g}}$, 
and $m_{\chi_2^{\pm}}  \cong m_{\chi_{3,4}^0} \gg m_{\chi_1^0}$. 
Corrections to these relations are $O(M_Z^2/\mu)$.
One also finds   that the light Higgs mass obeys $m_h\alt  120\ GeV$.
The lightest  neutralino $\chi_1^0$ is the CDM particle for almost all 
the parameter space.
 
Calculations of the detector event rates for the $\chi_1^0$ particles in the 
Milky Way incident on a terrestial nuclear target proceeds in two steps. 
One first calculates the relic density $\Omega_{\chi_1^0} h^2$ remaining after 
annihilation in the early universe \cite{7}.
Two regions occur here: for $m_{\chi_1^0}\alt 60\ GeV$ (or by scaling, 
$m_{\tilde{g}} \alt 450\ GeV$)
the annihilation is dominated by $h$ and $Z$ $s$-channel poles \cite{8}.
For higher $m_{\chi_1^0}$, the $t$-channel squark/slepton poles become dominant.
(These two regimes will show up below in detector event rates.)
One then restricts the SUSY parameter space so that the allowed window of 
$\Omega_{\chi_1^0}h^2$ is satisfied, as well as current accelerator SUSY bounds.
(These latter limit 
$A_t/m_0\agt 
 -0.5$
from the $t$-quark mass \cite{9} and eliminate most of the parameter space 
with $\mu<0$ from the 
$b\rightarrow s+\gamma$ decay data \cite{10}.)
One then calculates the expected terrestial detector event rates \cite{11} 
$R\ [events/kg\ d]$ for this restricted parameter space.

The $\mu$ parameter and $\tan\beta$ play an important role in determining $R$.
Thus large (small) $\tan\beta$ generally gives rise to large (small) $R$. 
Also as $\mu$ increases (decreases) one finds generally $R$ decreases (increases).
$m_0$ and $m_{\tilde{g}}$ play a more indirect role: since the early universe 
annihilation cross section decreases with $m_{\chi_1^0}$ and hence with 
$m_{\tilde{g}}$, the upper (lower) bounds on $\Omega_{\chi_1^0}h^2$ 
result in upper (lower) bounds on $m_{\chi_1^0}$ and $m_{\tilde{g}}$. 
Similarly, annihilation decreases with increasing $m_0$, generally restricting 
$m_0\alt 
 200\ GeV$ (so that the upper bound in $\Omega_{\chi_1^0}h^2$ is not violated)
except in the domain 
$m_{\chi_1^0}\alt 
 60\ GeV$ where $h$ and $Z$ poles produce such rapid annihilation 
that a large $m_0$ can be accomodated.

\vspace{.1in}

\noindent
{\it $\nu$CDM Model}.
We assume here that measurements of the cosmological parameters have yielded 
the central values of $\Omega_{\nu}=0.20$, $\Omega_B=0.05$ and $h=0.62$, 
(which are in accord with current estimates) and that $\Omega_{total}=1$.
This implies $\Omega_{\chi_1^0}=0.75$. 
The errors in which the Planck sattelite can measure the various quantities have 
been estimated in Refs.\cite{12,13} and we find from this
\begin{equation}
\Omega_{\chi_1^0}h^2 = 0.288 \pm 0.013
\eqnum{2} 
\label{eq:4}
\end{equation}
 This may be compared with the current estimate 
$0.1\le \Omega_{\chi_1^0}h^2 \le 0.4$ and shows the remarkable accuracy future 
sattelite measurements may be expected to obtain. 
Fig.~\ref{1} shows the maximum and minimum event rates for a xenon detector for universal 
soft breaking  with $0.275\le \Omega_{\chi_1^0}h^2 \le 0.301$, the 1 std window 
of Eq.\ (\ref{eq:4}). 
One may compare this with Fig.~\ref{2} of Ref.\cite{5} which uses the current estimate 
$0.1\le \Omega_{\chi_1^0}h^2 \le 0.4$, and shows the rise in event rates  
with increasing $m_{\chi_1^0}$ for $m_{\chi_1^0}\alt 
 60\ GeV$
($m_{\tilde{g}}\cong (7-8) m_{\chi_1^0} \cong 450\ GeV$) 
the $h$ and $Z$ pole dominated region in the early universe annihilation cross section, 
and the fall off for higher $m_{\chi_1^0}$.
Fig.~\ref{1} shows a narrowing of the predicted range of allowed event rates for 
$m_{\tilde{g}}\alt 
 450\ GeV$ (relative to Fig.2 of \cite{5}), 
and most striking the appearance of forbidden regions, i.e. gaps, in the allowed 
values of $m_{\tilde{g}}$ in the region $m_{\tilde{g}} \cong 500\ GeV$ and 
$m_{\tilde{g}} \cong 600\ GeV$. 
Fig.~\ref{2}, with $\delta_2=-1=-\delta_1$ shows the effects of non-universal soft 
breaking.
The event rates are reduced, 
(as expected from Eq.\ (\ref{eq:2}) since $\mu^2$ is increased) and the gap at 
$m_{\tilde{g}} \cong 500\ GeV$ is significantly widened. 
For the case $\delta_2=1=-\delta_1$, where $\mu^2$ is decreased, one finds 
the event rates are significantly increased i.e.
$10\agt \
 R\ (events/kg\ d)\agt 
 10^{-3}$, the upper bound being at the current sensitivity 
of NaI detectors \cite{14}.
However, for this case the gaps have disappeared, and there is now a lower bound of 
$m_{\tilde{g}}\agt 
 420\ GeV\ (m_{\chi_1^0}\agt 
 55\ GeV)$. 
In all cases one finds 
$m_{\tilde{g}}\alt 
 720\ GeV\ (m_{\chi_1^0}\alt 
 100\ GeV)$ 
in order that the upper bound, $\Omega_{\chi_1^0} h^2 \le 0.301$, 
is obeyed.

As commented above one expects violations of the scaling relations 
 of size $O(M_Z^2/\mu)$.
For $\delta_2=-1=-\delta_1$, where $\mu^2$ is increased, the effects are reduced, 
but can become quite significant for $\delta_2=1=-\delta_1$ in the region 
$m_{\chi_1^0} \alt
 60\ GeV$.
The effects of non-universal soft-breaking are suppressed, however, for 
$m_{\chi_1^0}\agt 
 60\ GeV\ (m_{\tilde{g}}\agt 
 450\ GeV)$  since the universal 
contributions in  Eq.\ (\ref{eq:2}) dominate.

\vspace{.1in}

\noindent
{\it $\Lambda$CDM Model}. 
We consider as a second example with $\Omega_{total}=1$, the possibility 
that the new measurements show the existence of a large cosmological constant 
with $\Omega_{\Lambda}=0.55$. 
In addition we assume here baryonic matter with $\Omega_B=0.05$ and $h=0.62$. 
This would imply $\Omega_{\chi_1^0}=0.40$. 
(The above choice for $\Omega_{\Lambda}$ represents the current upper limit 
\cite{15}, while the value for $\Omega_{\chi_1^0}$ is approximately what is 
actually observed from large galactic clusters.)
Using the estimated errors expected for the Planck sattelite \cite{12} one finds 
$\Omega_{\chi_1^0} h^2 = 0.154 \pm 0.017$. 
Fig.~\ref{3} shows the maximum and minimum event rates for a xenon detector for three 
examples of $\delta_i$.
We again see a reduction of event rates for $\delta_2=-1=-\delta_1$ and the 
enhancement for $\delta_2=1=-\delta_1$.
This time one finds a low upper bound on $m_{\tilde{g}}$ of 
$m_{\tilde{g}}\alt 
 520\ GeV\ (m_{\chi_1^0} \alt
 70\ GeV)$ due to the relatively 
low value of $\Omega_{\chi_1^0} h^2$. 
For the case $\delta_2=1=-\delta_1$ the event rates for $\mu>0$ are relatively 
high, i.e.  $(3\times 10^{-3} - 1.0)\ events/kg\ d$, and there is again a minimum 
gluino mass for this case: $m_{\tilde{g}} \ge 400\ GeV\ (m_{\chi_1^0} \ge 50\ GeV)$. 
Thus  the choice $\delta_2=1=-\delta_1$ sharply restricts the allowed gluino 
and neutralino mass range. 

As mentioned above, there exists a correlation between large (small) values of 
detector event rate  
$R$ and small (large) values of the branching ratio $B(b\rightarrow s+\gamma)$. 
This is exhibited in Fig.~\ref{4} which shows a scatter plot of values of $R$ and $B$. 
The current experimental value is \cite{16} $B= (2.32\pm 0.67)\times 10^{-4}$ 
while the Standard Model (SM) with NLO corrections predicts \cite{17}  
$B= (3.51\pm 0.31)\times 10^{-4}$. 
One sees that almost all points with $B < 2.8\times 10^{-4}$ (i.e. $>2$ std 
below the SM 
prediction) have $R>0.1\ events/kg\ d$, while almost all points with 
$B > 3.0\times 10^{-4}$ (i.e. $>1$ std above the experimental value) have 
$R<0.1\ events/kg\ d$.
Future CLEO data should significantly reduce the experimental error in $B$,
and hence may influence the expected size of $R$.

The domain $m_{\tilde{g}}\alt 400\ GeV$ corresponds to the 
$Z$ and $h$ pole dominated region in the early universe annihilation cross 
section and hence $m_0$ can be large 
so that squark and slepton masses can be large. 
For $m_{\tilde{g}} \agt 420\ GeV$
$m_0$ is generally small ($m_0 \alt 100\ GeV$)
so that the upper bound on $\Omega_{\chi_1^0} h^2$ not be violated, and so in 
this domain  the lightest slepton, $\tilde{e}_R$, could have a mass as low as 
$(85-90)\ GeV$,  at the edge of detectability of LEP190. 
The squarks are also relatively light in this domain i.e. $\alt (400-500)\ GeV$.   

\vspace{.1in}

\noindent
{\it Conclusions}.
We have considered in this letter what the future precision measurements of the 
basic cosmological parameters  will imply for accelerator searches for SUSY.
In the $\nu$CDM model the gluino mass obeys 
$m_{\tilde{g}} \alt
 720\ GeV$, and more significantly there can be forbidden regions 
at  $m_{\tilde{g}} \approx 500\ GeV$ and $m_{\tilde{g}} \approx 600\ GeV$. 
Violations of the gaugino scaling laws can become significant for 
 $m_{\chi_1^0}\alt 
 65\ GeV$.
For the $\Lambda$CDM model one finds that 
$m_{\tilde{g}}\alt 520\ GeV$ ($m_{\chi_1^0}\alt 70\ GeV$),  
and for  $m_{\tilde{g}}\agt 420\ GeV$  ($m_{\chi_1^0}\agt 55\ GeV$)  
one has relatively light squarks and can have a light $\tilde{e}_R$ selectron. 
In both models there is a general correlation between large dark matter detector 
event rates and small $b\rightarrow s+\gamma$ branching ratio 
(e.g. $B(b\rightarrow s+\gamma)$ lying significantly below the SM predictions).
In general both models are sensitive to non-universal SUSY soft breaking.
The analysis given here shows that accurate determinations of the cosmological 
parameters will significantly constrain many aspects of what would be expected 
at accelerators. 
While detection of SUSY dark matter is only one supersymmetric phenomenon, one 
might expect to learn from this two items: the magnitude of the event rate and the 
mass  of the $\chi_1^0$. 
These could greatly further constrain what would be expected at accelerators. 

This work was supported in part by National Science Foundation Grants 
PHY-96020274 and PHY-9722090.

\begin{figure}
\caption{
Maximum and minimum event rates for $\mu>0$ for a xenon detector as a 
function of $m_{\tilde{g}}$ for the 1 std range of Eq.\ (\protect\ref{eq:4})
of the $\nu$CDM model with $\delta_i=0$.
}
\label{1}
\end{figure}

\begin{figure}
\caption{
Same as Fig.~\protect\ref{1} with $\delta_2=-1=-\delta_1$, $\delta_3=0=\delta_4$.
}
\label{2}
\end{figure}

\begin{figure}
\caption{
Maximum and minimum event rates for a xenon detector for the 1 std band 
of the $\Lambda$CDM model with $\delta_1=0=\delta_2$ (solid), 
$\delta_2=-1=-\delta_1$ (dotted), 
$\delta_2=1=-\delta_1$ (dashed) for $\delta_3=0=\delta_4$.
The high mass discrete points are for $\delta_2=1=-\delta_1$.
}
\label{3}
\end{figure}

\begin{figure}
\caption{
Scatter plot of  $R$ vs $B(b\rightarrow s+\gamma)$ for the 
$\Lambda$CDM model, $\mu>0$ and $\delta_i=0$ for the 1 std range.
}
\label{4}
\end{figure}

\end {document}
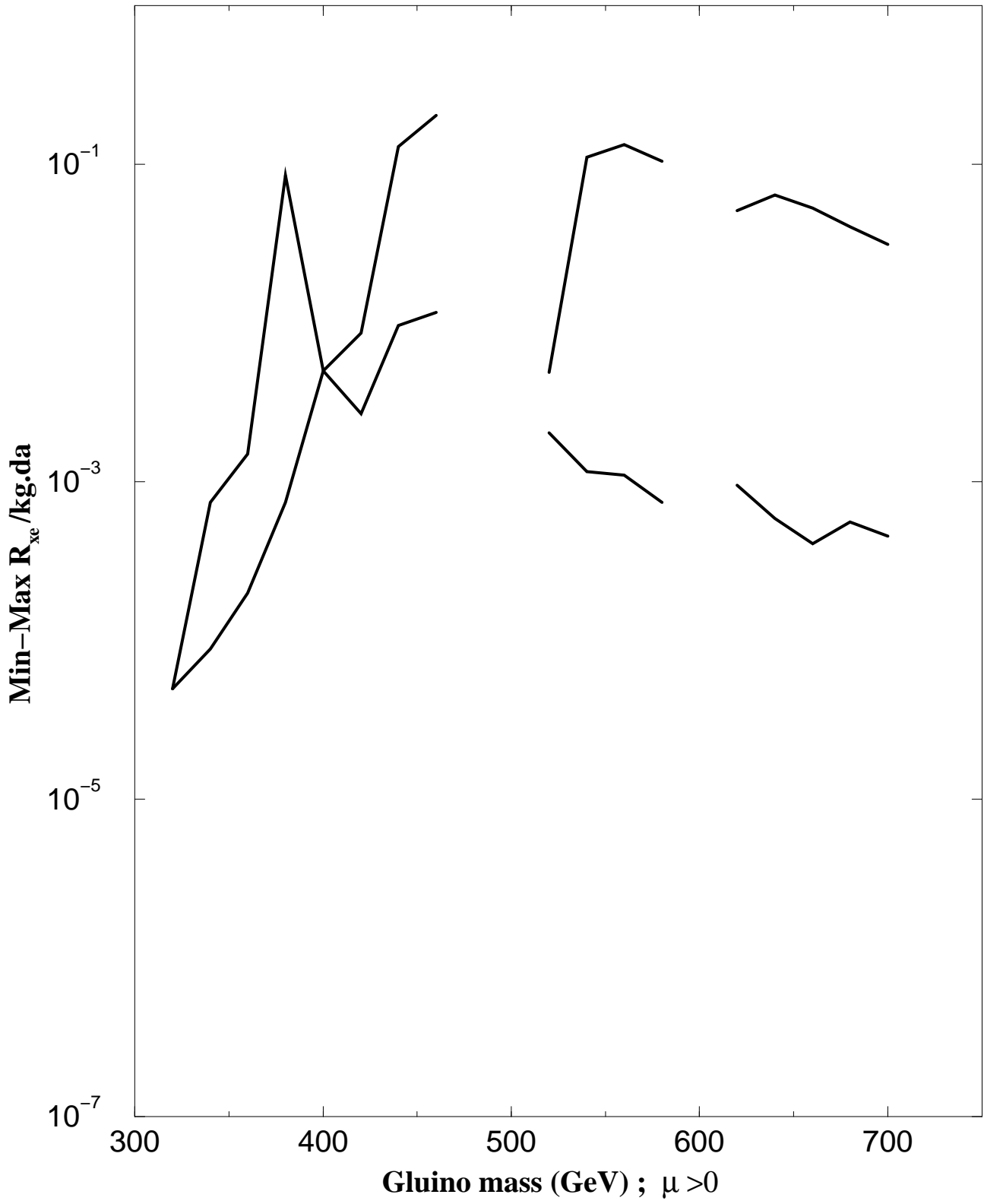
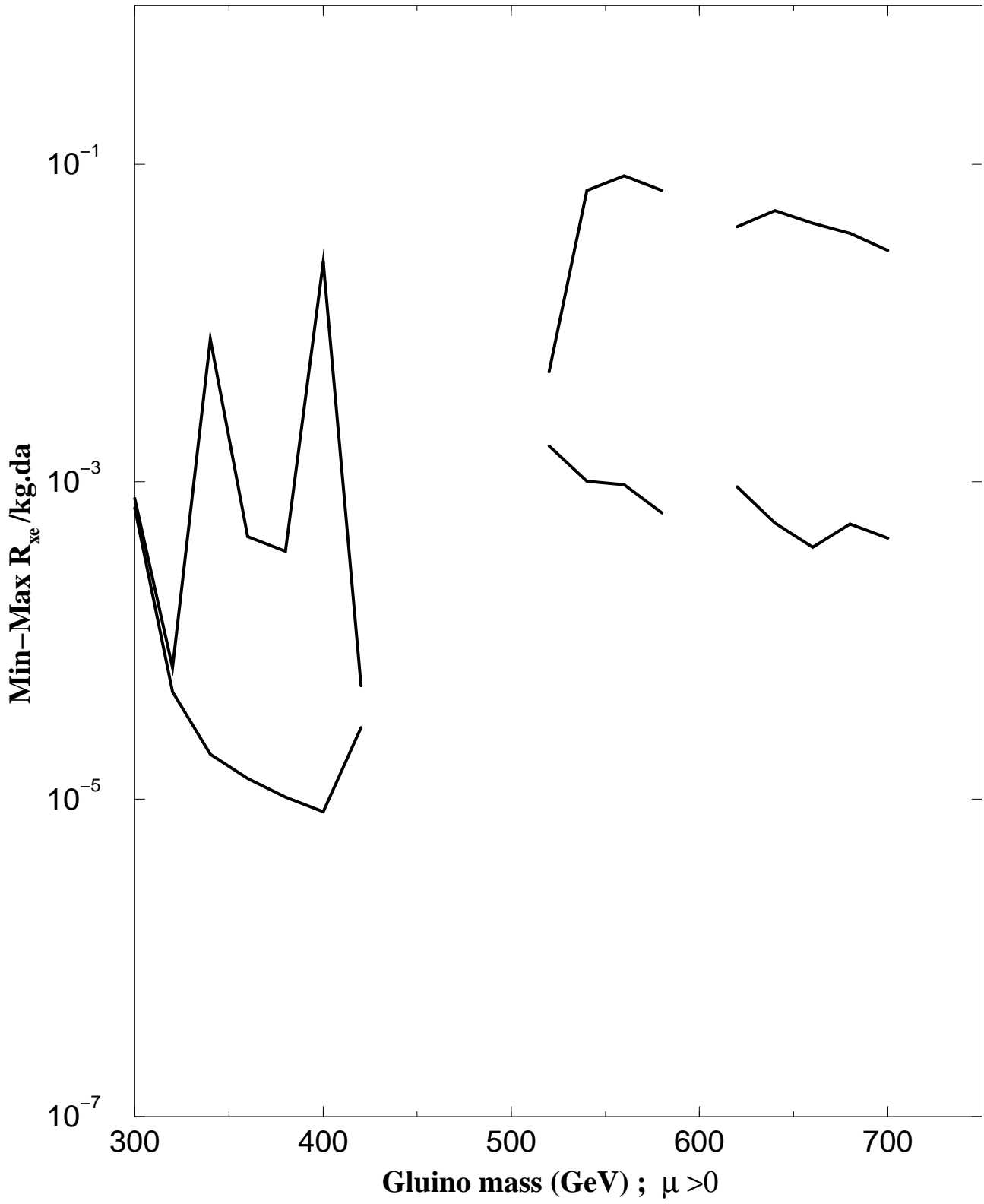
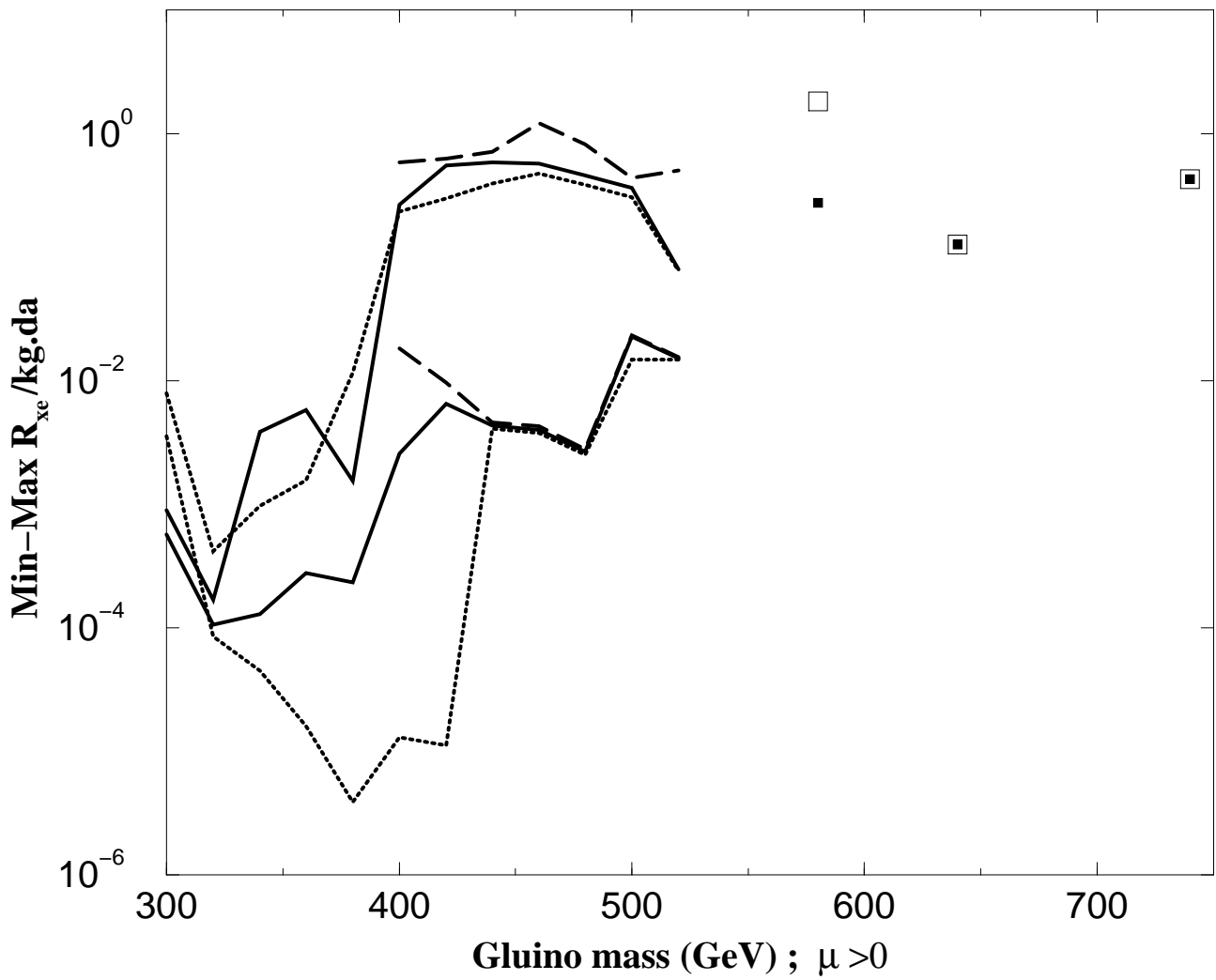
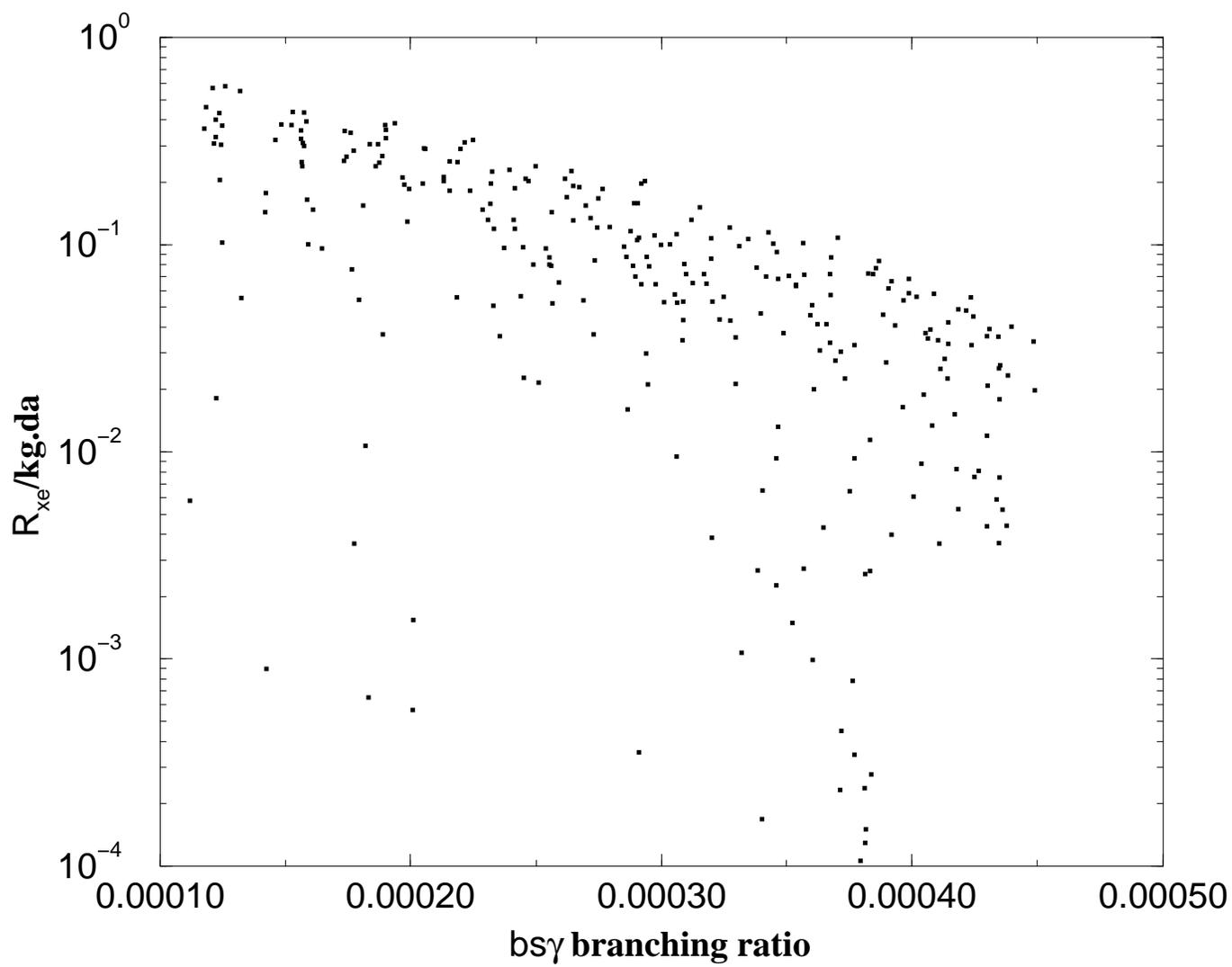
